\documentclass{JINST}
\pdfoutput=1
\title{Micromegas detector developments for Dark Matter directional detection with MIMAC}

\author{F.J.~Iguaz$^a$, D.~Atti\'e$^a$, D.~Calvet$^a$, P.~Colas$^a$, F.~Druillole$^a$, E.~Ferrer-Ribas$^a$\thanks{Corresponding
author.}, I.~Giomataris$^a$, J.P.~Mols$^a$, J.~Pancin$^b$, T.~Papaevangelou$^a$, J.~Billard$^c$, G.~Bosson$^c$, J.L.~Bouly$^c$, O.~Bourrion$^c$, Ch.~Fourel$^c$, C.~Grignon$^c$, O.~Guillaudin$^c$, F.~Mayet$^c$, J.P.~Richer$^c$, D.~Santos$^c$, C.~Golabek$^d$ and 
L.~Lebreton$^d$\\
\llap{$^a$}CEA/DSM/IRFU,\\
  CEA, 
  91191 Gif sur Yvette,  France\\  
\llap{$^b$}GANIL,\\
 Bvd H. Becquerel, Caen, France\\
\llap{$^c$}LPSC,\\
  Universite Joseph Fourier Grenoble 1, CNRS/IN2P3, Institut Polytechnique de
Grenoble, France\\
\llap{$^d$}LMDN,\\
  LMDN, IRSN Cadarache, 13115 Saint-Paul-Lez-Durance, France\\
  E-mail: \email{esther.ferrer.ribas@cea.fr}}

\abstract{The aim of the MIMAC project is to detect non-baryonic Dark Matter with a directional  TPC using a high 
precision Micromegas readout plane. We will describe in detail the recent developments done with bulk Micromegas detectors as well as the characterisation measurements performed in an Argon(95\%)-Isobutane(5\%) mixture. Track measurements with alpha particles  will be shown.}


\begin{document}
\section{Introduction}
      
The MIMAC (MIcro TPC MAtrix of Chambers) collaboration\cite{mimac} aims at building a directional Dark Matter detector composed of a matrix of Micromegas\cite{mm96} detectors. The MIMAC project is designed to measure both 3D track and ionization energy of recoiling nuclei, thus leading to the possibility to achieve directional dark Matter detection\cite{spergel}. It is indeed a promising search strategy of galactic Weakly Interacting Massive Particles (WIMPs) and several projects of detector are being developed for this goal\cite{review}. Recent studies have shown that a low exposure CF4 directional detector could lead either to a competitive exclusion\cite{billard1}, a high significance discovery\cite{billard2}, or even an identification of Dark Matter\cite{billard3}, depending on the value of the WIMP-nucleon axial cross section.

Gaseous detectors present the advantage of being able to reconstruct the track of the nuclear recoil and to access both 
the energy and the track properties. Micropattern gaseous detectors are particularly suited to reconstruct low energy (few keV) 
recoil tracks  of a few mm  length due to their very good granularity, good spatial and energy resolution and low threshold. Micromegas detectors have shown these qualities in different environments \cite{cast1,cast2,tpchp,T2Kdetect}. In particular thanks to the new manufacturing techniques, namely bulk\cite{bulk} and microbulk\cite{microbulk}, where the amplification region is produced as a single entity. In bulk Micromegas a woven mesh is laminated on a printed circuit board covered by a photoimageable film and the pillars are made by a photochemical technique with insulation through a grid. This technique can be transfered to industry allowing the production of large, robust inexpensive detector modules.

This paper describes the developments done with bulk Micromegas detectors in order to show  the feasability of a large TPC (Time Projection Chamber) for directional detection. Section~\ref{sec:concept} describes briefly the strategy of the MIMAC project. In section~\ref{sec:design} we  discuss in detail the design of the first prototype detector of $10\times10$\,$\mathrm{cm}^2$. The experimental set-up used for the characterisation in the laboratory  is presented in section \ref{sec:setup} and the results are given in section~\ref{sec:characterisation}. Section~\ref{sec:results} is devoted to the the results obtained for the reconstruction of tracks with alpha particles. Finally, the conclusions and the perspectives are discussed in section~\ref{sec:conclusions}.

\section{The MIMAC concept}\label{sec:concept}
The nuclear recoil produced by a WIMP in the TPC  produces electron-ion pairs in the conversion gap of the Micromegas detector that drift towards the amplification gap (128$\,\mu\mathrm{m}$ or 256$\,\mu\mathrm{m}$ in this case) producing an avalanche that will induce signals in the x-y anode and in the mesh. The third dimension z of the recoil is reconstructed by a dedicated  self-triggered electronics specifically designed for this project~\cite{mimacelec,DAQ} that is able to perform anode sampling at a frequency of 40 MHz. 
The concept had already been tested with a prototype detector of small size 3$\times$ 3 cm$^2$ and with the first version of the electronics\cite{mimac,Quenching}.

The aim of building a detector of 10 $\times$ 10 cm$^2$ was to validate the feasability of a large TPC for directional detection with a realistic size prototype. The design of the bulk Micromegas was guided by the requirements on the granularity of the anode as well as by the operation conditions. Simulation studies showed that the granularity of the readout plane needed strips of 200\,$\mu\mathrm{m}$ size. The MIMAC project is expected to work with two different regimes: high pressure (up to 3~bar) in order to improve the probability to have WIMP interactions and low pressure (down to about 100  of~mbar) to get the directionality information of the WIMPs. The design of the bulk Micromegas end cap should take into account these requirements. From the beginning of the design, it was known that the detector would be first validated in the laboratory with the T2K electronics~\cite{T2Kelec1,T2Kelec2} before the final conclusive test with the specifically designed MIMAC electronics~\cite{mimacelec,DAQ}. Special care was taken in the design to have a portable system.

\section{Design of the Bulk Micromegas detector: 10 $\times$ 10 cm$^2$}\label{sec:design}
\begin{figure}[htb!]
\centering%
\begin{tabular}{cc}
\includegraphics[width=0.5\textwidth]{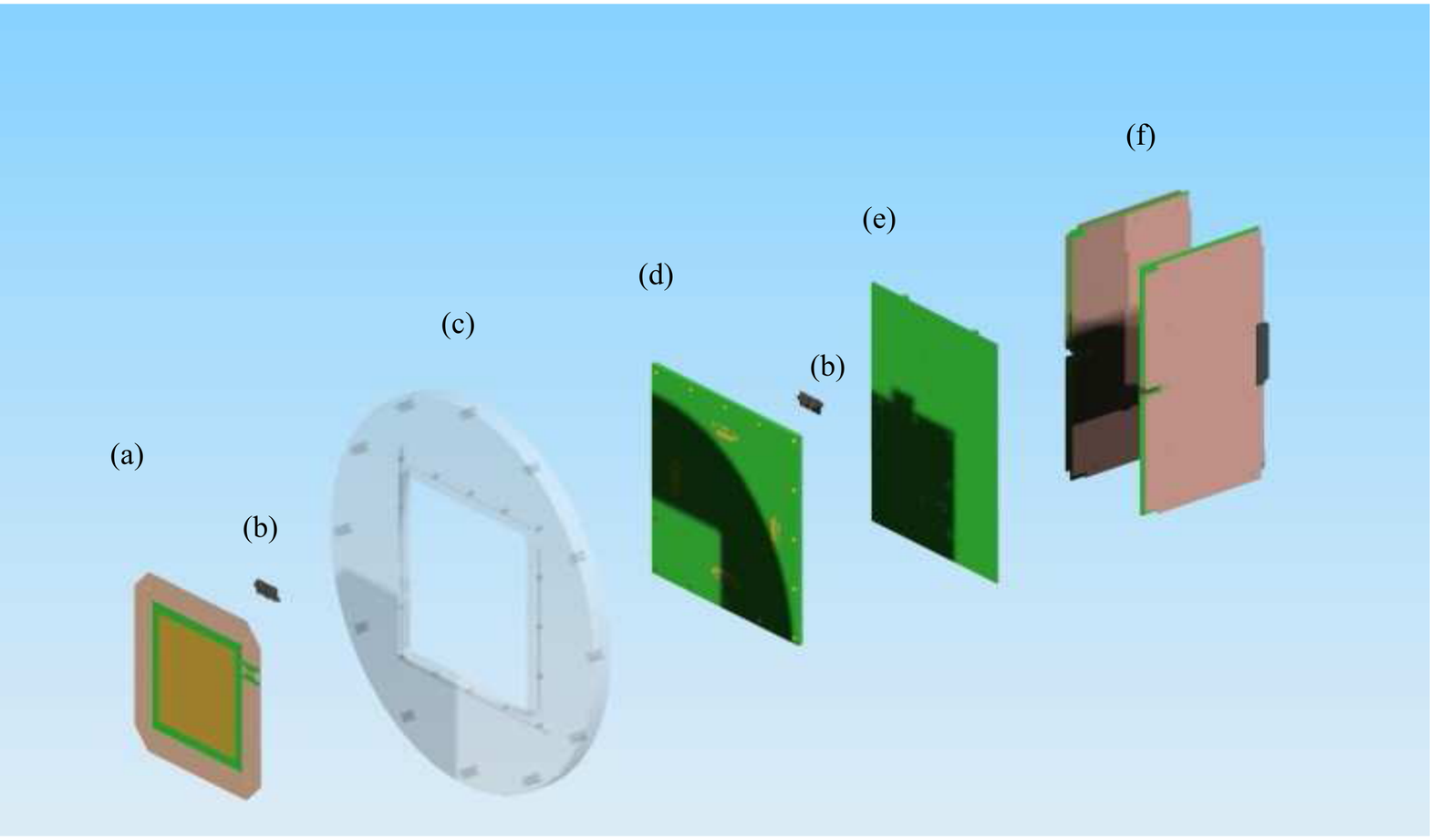} &
\includegraphics[width=0.5\textwidth]{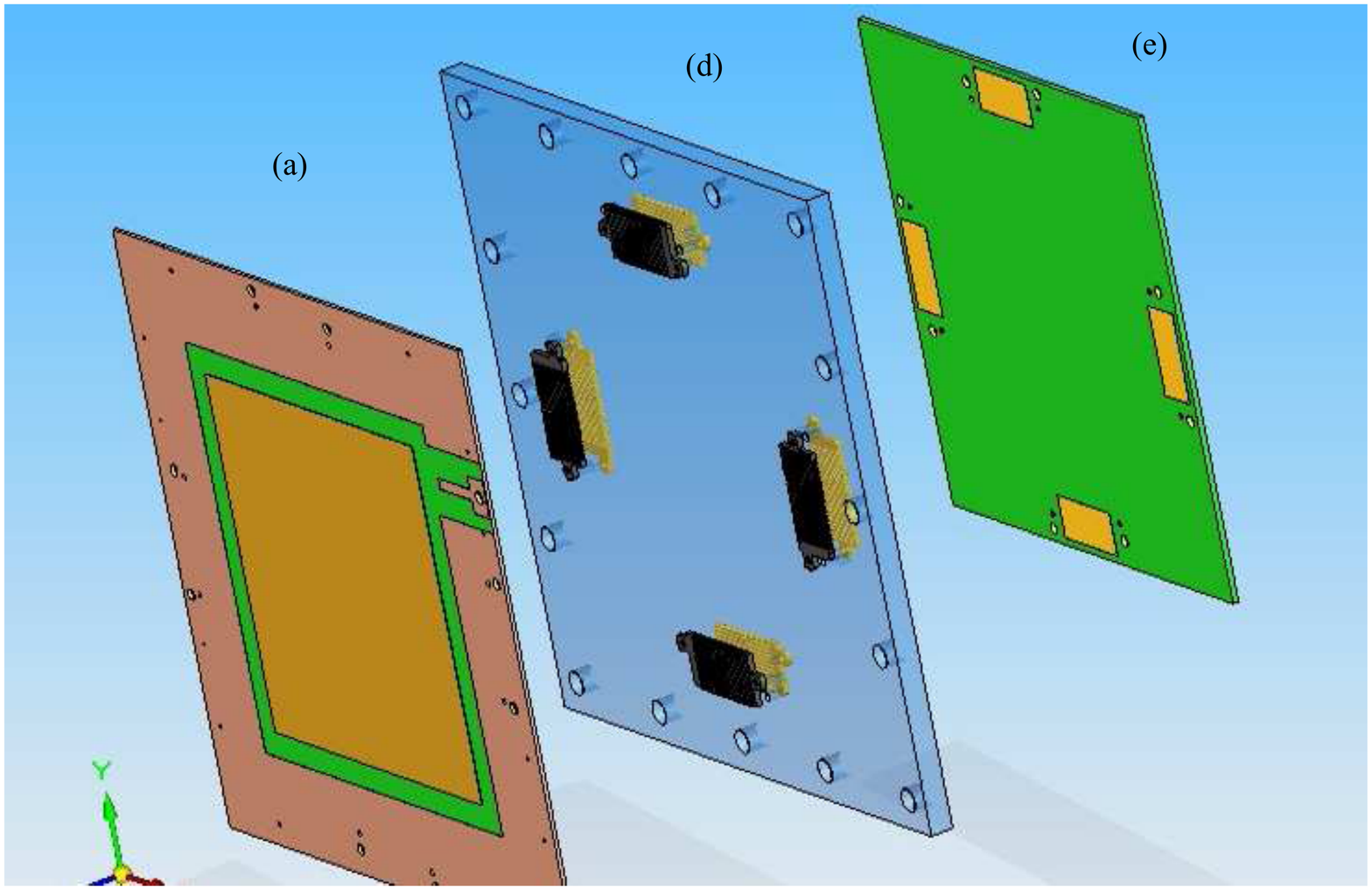}
\end{tabular}
\caption{Left: Sketch of the general assembly constituting the readout plane and the interface to electronics. Left: Zoom of the Readout PCB (\emph{a} in the image), Leak tight PCB (\emph{d}) and the Interface PCB (\emph{e}). }
\label{fig:assembly}
\end{figure}

In order to have a detector that can stand operation at low and high pressure the design relies on the idea of assembling a 
leak-tight read-out plane on a 2\,cm aluminium cap.  A general sketch of the mechanical assembly is given in figure~\ref{fig:assembly}.

\begin{figure}[htb!]
\centering%
\begin{tabular}{cc}
\includegraphics[width=0.45\textwidth]{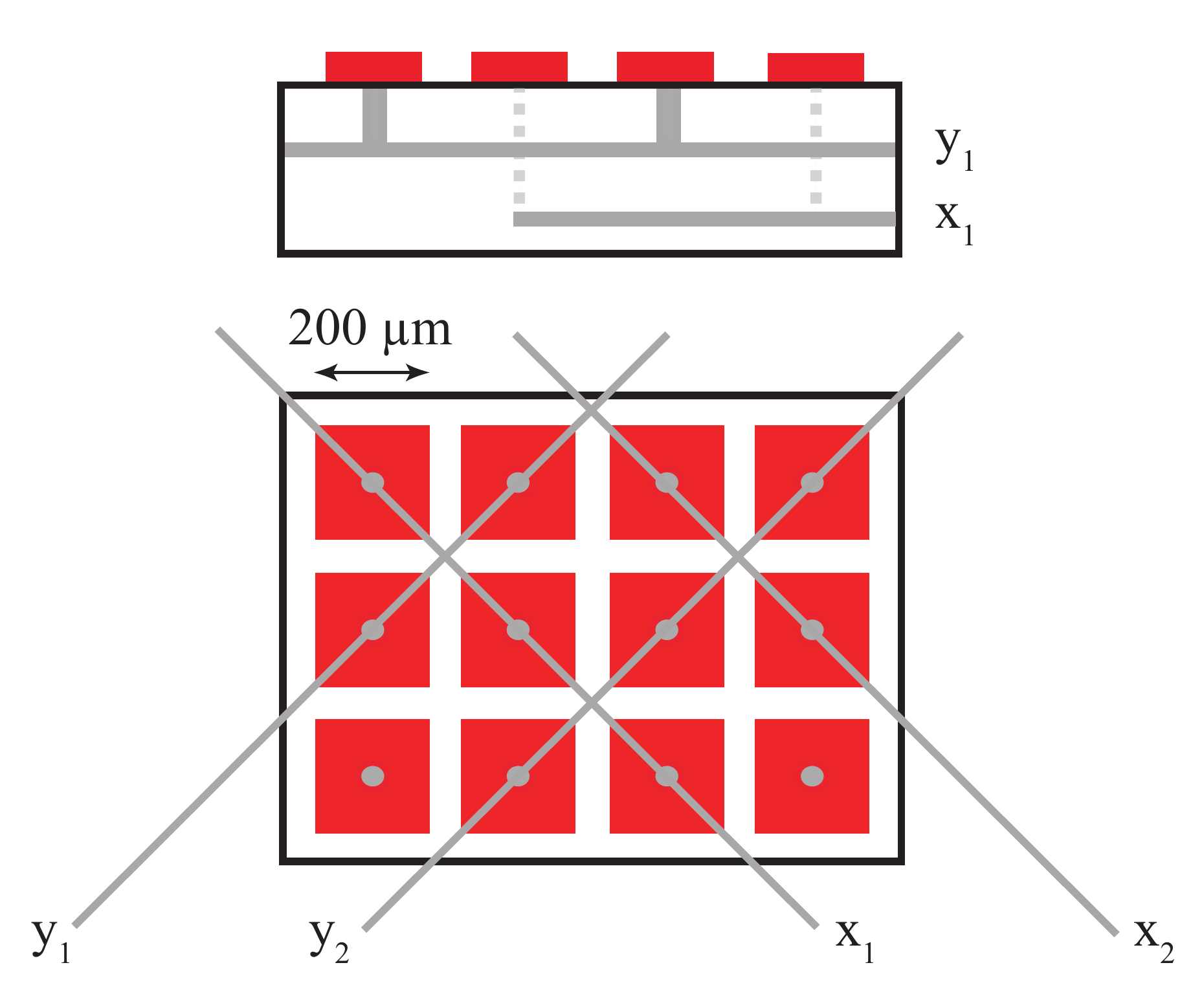} &
\includegraphics[width=0.45\textwidth]{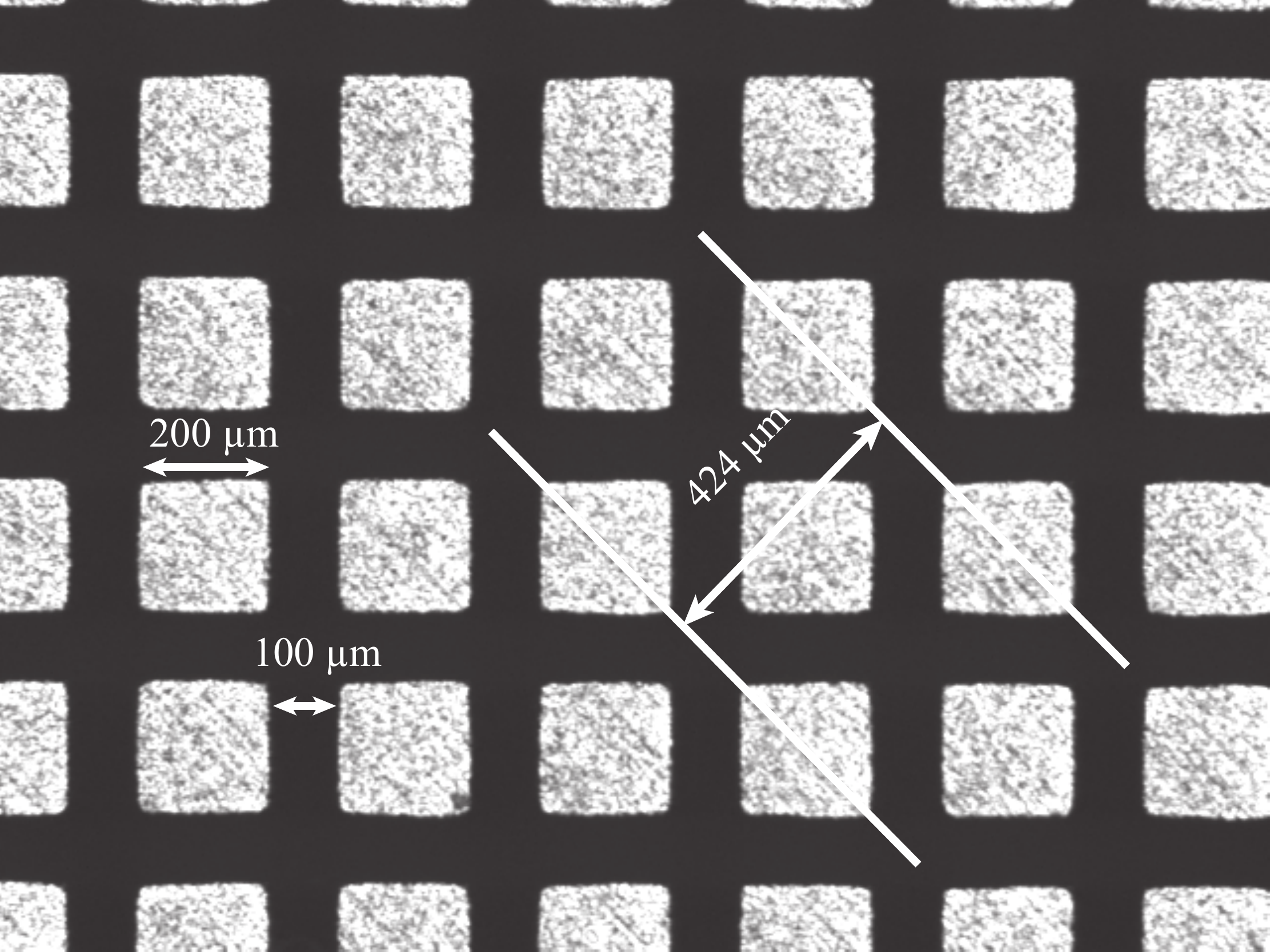}
\end{tabular}
\caption{Left: Sketch of the 2D readout used. Right: Microscope photograph of the 2D readout.}
\label{fig:2d}
\end{figure}

\begin{figure}[htb!]
\centering%
\begin{tabular}{c}
\includegraphics[width=0.5\textwidth]{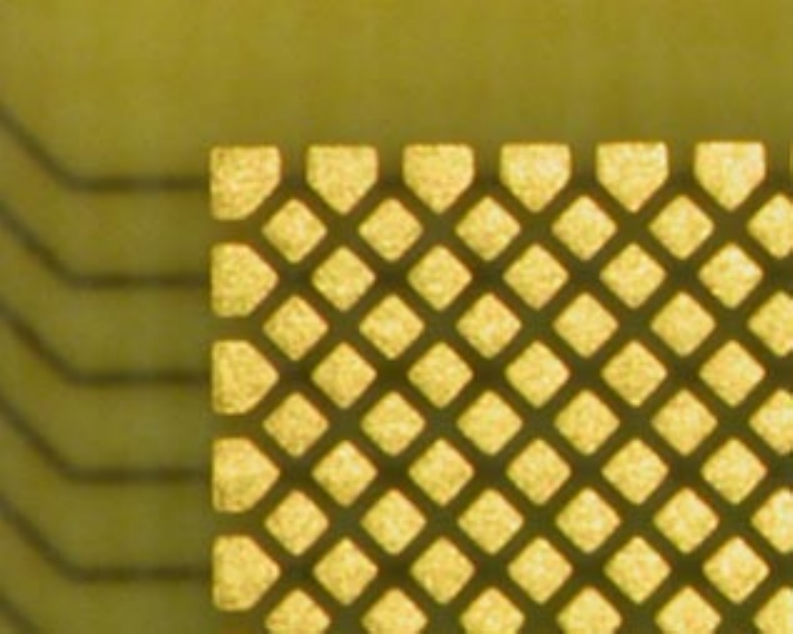} 
\end{tabular}
\caption{In this picture the pixel and the readout strips can be seen. 
The pixels are at 45$^\mathrm{o}$ with respect to the readout strips (observed by transparency).}
\label{fig:2dolivier}
\end{figure}

The bulk Micromegas is on a Printed Circuit Board (PCB), called \emph{Readout PCB}, of 1.6\,mm thickness (\emph{a} in figure~\ref{fig:assembly}).  The active surface is of 10.8$\times$ 10.8\,cm$^2$ with 256 strips per direction. The charge collection strips make-up an X--Y structure out of electrically connected pads in the diagonal direction through metallized holes as can be seen in figures~\ref{fig:2d} and \ref{fig:2dolivier}. This readout strategy  reduces the number of channels with a fine granularity covering a large anode surface. The pads are 200\,$\mu\mathrm{m}$ large with an isolation of  100~$\mu\mathrm{m}$ resulting into a strip pitch of 424~$\mu\mathrm{m}$. The quality of the surface of the readout plane can be observed in figure~\ref{fig:2d}. The 100\,$\mu\mathrm{m}$ diameter metallized holes have been fully filled yielding a completely uniform surface.  This fact is a prerequisite to  obtain a uniform performance of a bulk Micromegas detector.

The strips signals are rooted into 4 connectors prints  at the sides of the Readout PCB. The Readout PCB is screwed on a thick 0.5\,cm PCB, called \emph{Leak Tight PCB} (\emph{d} in figure~\ref{fig:assembly}), that will ensure the leak tightness of the system. The Leak Tight PCB is constituted of various layers of FR4 with blind metallized vias in the inner layer. This piece is then screwed on a 2\,cm thick aluminium cap that constitutes the bottom of the TPC (\emph{c} in figure~\ref{fig:assembly}). The signal connections from one board to another are done by means of SAMTEC connectors (GFZ 200 points) that are placed and screwed between the two boards (\emph{b} in figure~\ref{fig:assembly} left). On the outside of the vessel an \emph{Interface card} distributes the signals to the desired electronics (\emph{e} in figure~\ref{fig:assembly}). 
Two versions of this card exists: one dedicated to the laboratory set-up and a second one for the final MIMAC electronics.

This design offers several advantages: a simple, compact and leak-tight way for the signal connections and a versatility for two different types of electronics. 
Bulk Micromegas with two different amplification gaps were produced in order to choose the best gap for different running pressure conditions. Characterisations tests described in section~\ref{sec:characterisation} 
concern three readout planes with a gap of 128\,$\mu$m and two with a gap of 256\,$\mu$m.

\section{Experimental set up}\label{sec:setup}
\begin{figure}[htb!]
\centering%
\includegraphics[width=0.8\textwidth]{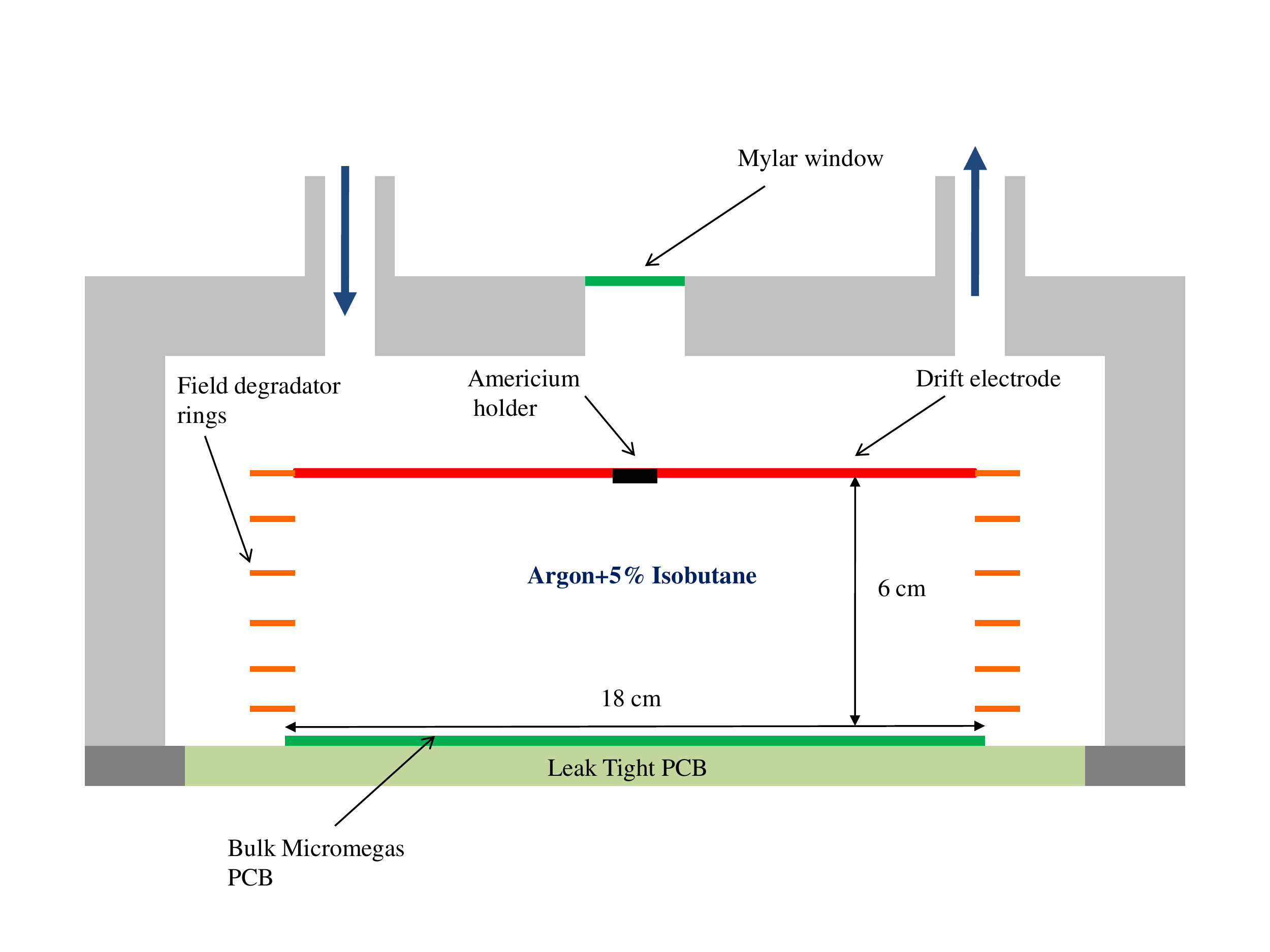} 
\caption{Schematic drawing of the set-up.}
\label{fig:setup}
\end{figure}

A dedicated vessel was built as shown in figure~\ref{fig:setup}. It consists of two aluminium caps
screwed to create the TPC. In the top cap, an iso-KF25 valve is used
for pumping in order to reduce the outgassing from the inner walls. This cap is  also equipped with  two
outlets for gas circulation and four SHV electrical connections. The bottom cap with the bulk Micromegas
has been described in section~\ref{sec:design}.

To uniformise the drift field in the conversion volume of 6\,cm of height, the vessel is equipped with a field shape degrador. It consists of 5 copper squared rings
and a copper plate, separated by a distance of 1\,cm by four peek columns.
The plate and the rings are electrically connected via resistors of 33\,M$\Omega$. The
voltage of the last ring is fixed with a variable resistor located outside the vessel.
The drift electrode has been designed to accommodate an $^{241}$Am source. The
source in its holder is screwed to the cathode plate. For illumination with X-ray and gamma sources 
a thin mylar window has been foreseen at the top of the vessel.

\section{Characterisation measurements and results}\label{sec:characterisation}

After the installation of the detector in the dedicated vessel, the TPC was closed and pumped to reduce the outgassing of the walls. A flow of 10 l/h of an  Argon(95\%)-Isobutane(5\%) mixture was circulated for some hours before starting the measurements. This gas was used for the characterisation measurements eventhough the chosen gas for dark matter search will be CF$_4$. Indeed most of Micromegas detectors are first characterised in Argon mixtures where their performance can be compared easily.  

The strips were connected to ground and the readout was illuminated by an iron $^{55}$Fe source (X-rays of 5.9 keV) located on the TPC window. The mesh voltage was typically varied from 300 to 450\,V for detectors with a gap of 128~$\mu$m and between 470 to 600\,V for those with a gap of 256\,$\mu$m. The drift voltage was changed from 300 to 3000\,V. Mesh and drift voltages were powered independently by CAEN N126 and CAEN N471A modules respectively. The signal induced in the mesh was read out by an ORTEC 142C preamplifier, whose output was fed into an ORTEC 472A Spectroscopy amplifier and subsequently into a multi-channel analyzer AMPTEK MCA-8000A for spectra building. Each spectrum contained at least $5 \times 10^4$ events and was fitted to get the peak position and the energy resolution.

\begin{figure}[htb!]
\centering
\includegraphics[width=100mm]{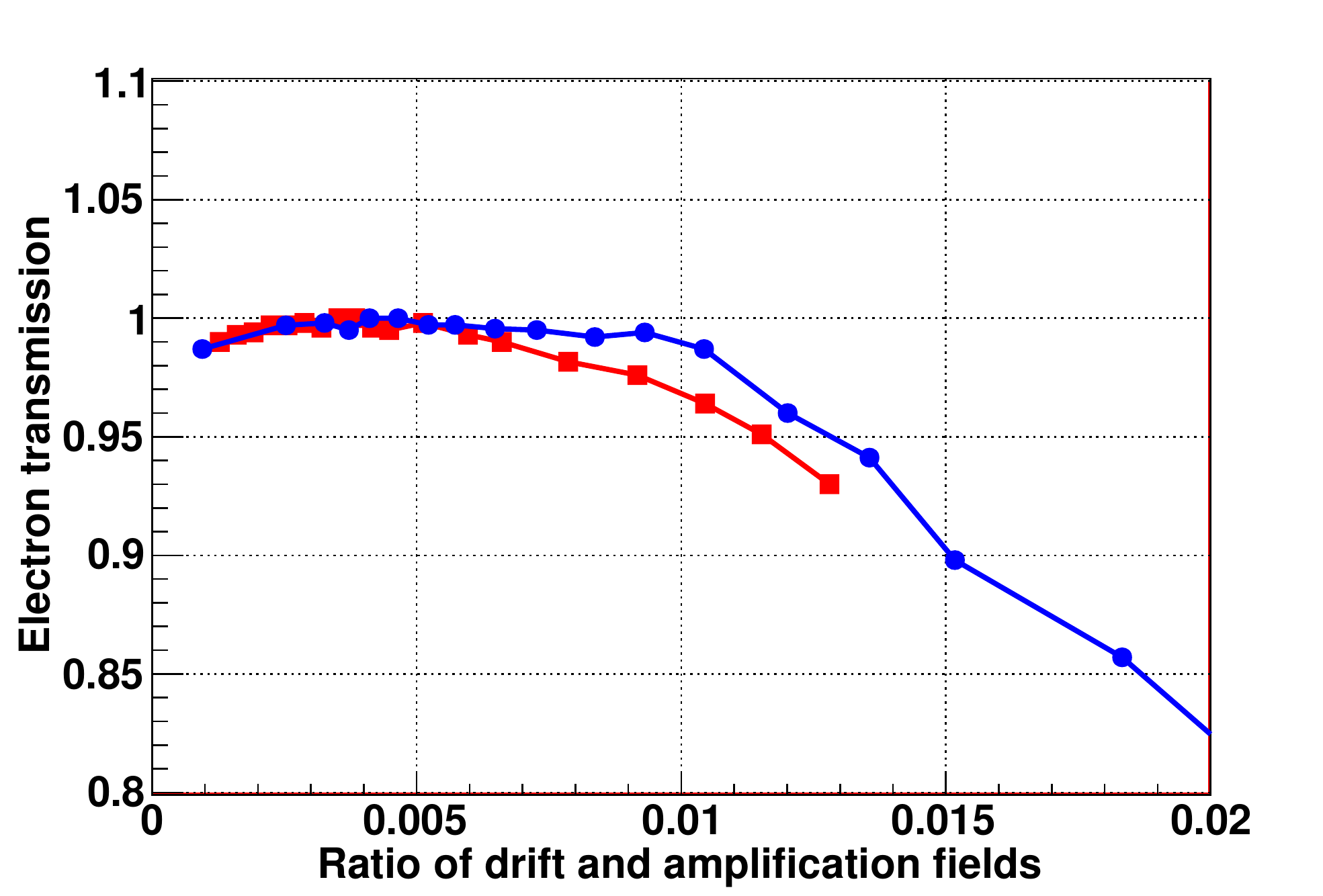}
\caption{ Dependence of the electron mesh relative maximum transmission with the ratio of the drift and amplification fields for the readouts with a 128\,$\mu$m (red squared line) and a 256\,$\mu$m-thick amplification gap (blue circled line).}
\label{fig:Trans10x10}
\end{figure}

In order to obtain the electron transmission curve, the drift voltage was varied for a fixed mesh voltage. Figure \ref{fig:Trans10x10} shows the typical plateau for Micromegas readout planes where the maximum electron transmission is obtained for a ratio of drift and amplification fields lower than 0.01. For ratios over this value, the mesh stops being transparent for the primary electrons generated in the conversion volume and both the gain and the energy resolution deteriorate for high drift fields. As shown in figure \ref{fig:Trans10x10}, readouts with a 256$~\mu$m-thick gap have a slightly larger plateau than those with a 128\,$\mu$m-thick gap. No other significant difference was observed among the tested readouts.

\begin{figure}[htb!]
\centering
\includegraphics[width=100mm]{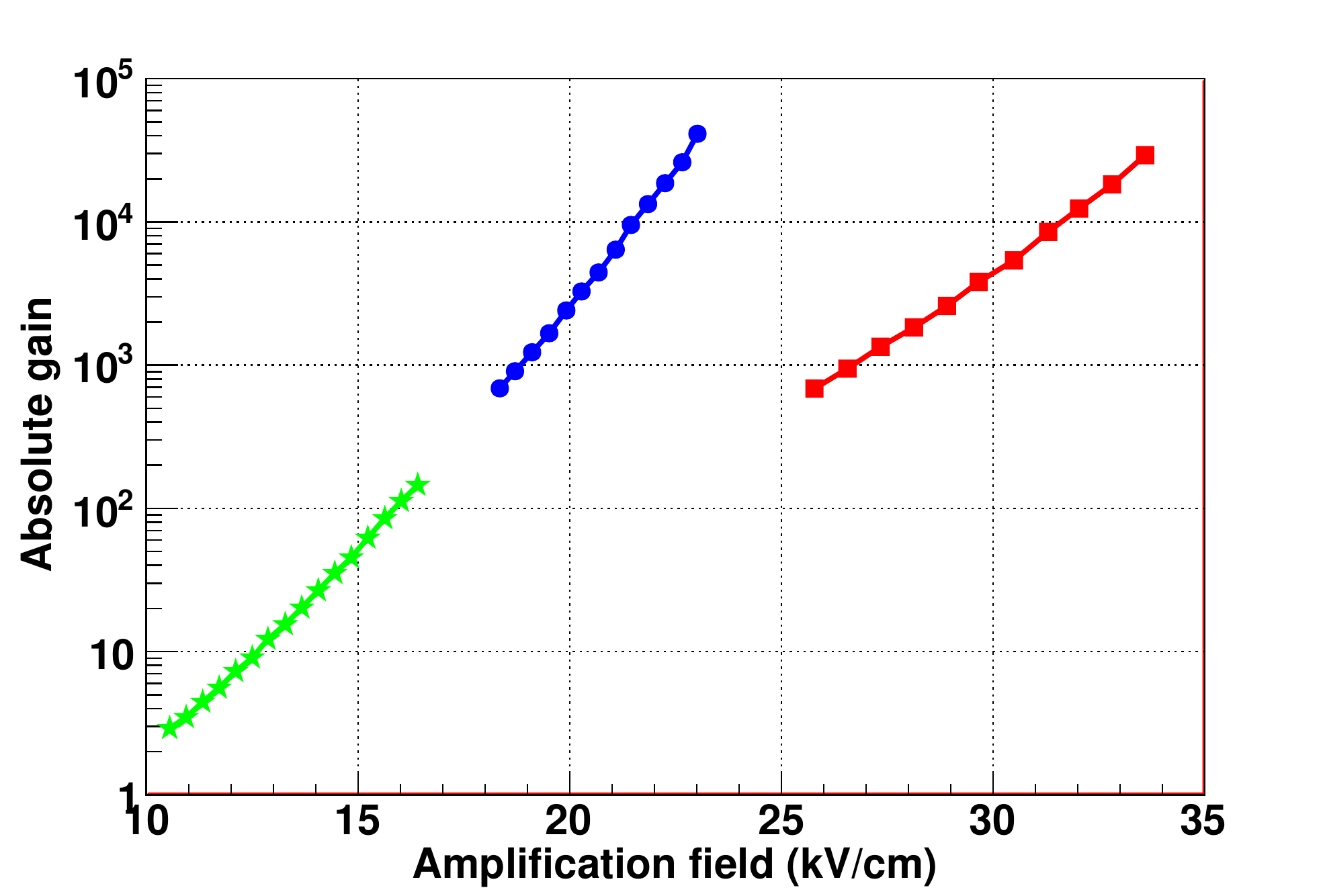}
\caption{The absolute gain as a function of the amplification field is shown for the readouts with a 128\,$\mu$m (red squared line) and 256\,$\mu$m-thick amplification gap (blue circled line) obtained with a iron $^{55}$Fe source using an Argon(95\%)-Isobutane(5\%) mixture at 1\,bar. The extension at low gain (green 
star line) for the
256\,$\mu$m readout was obtained with an alpha source. The maximum gain was obtained before the spark limit.}
\label{fig:Gain10x10alpha}
\end{figure}

After this first test, the ratio of drift and amplification fields was fixed in the region where the mesh showed the maximum electron transmission and the mesh voltage was varied to obtain the gain curves, shown in figure \ref{fig:Gain10x10alpha}. The tested readouts  reach gains greater than $2 \times 10^4$ before the spark limit for both amplification gaps.

\begin{figure}[htb!]
\centering
\includegraphics[width=100mm]{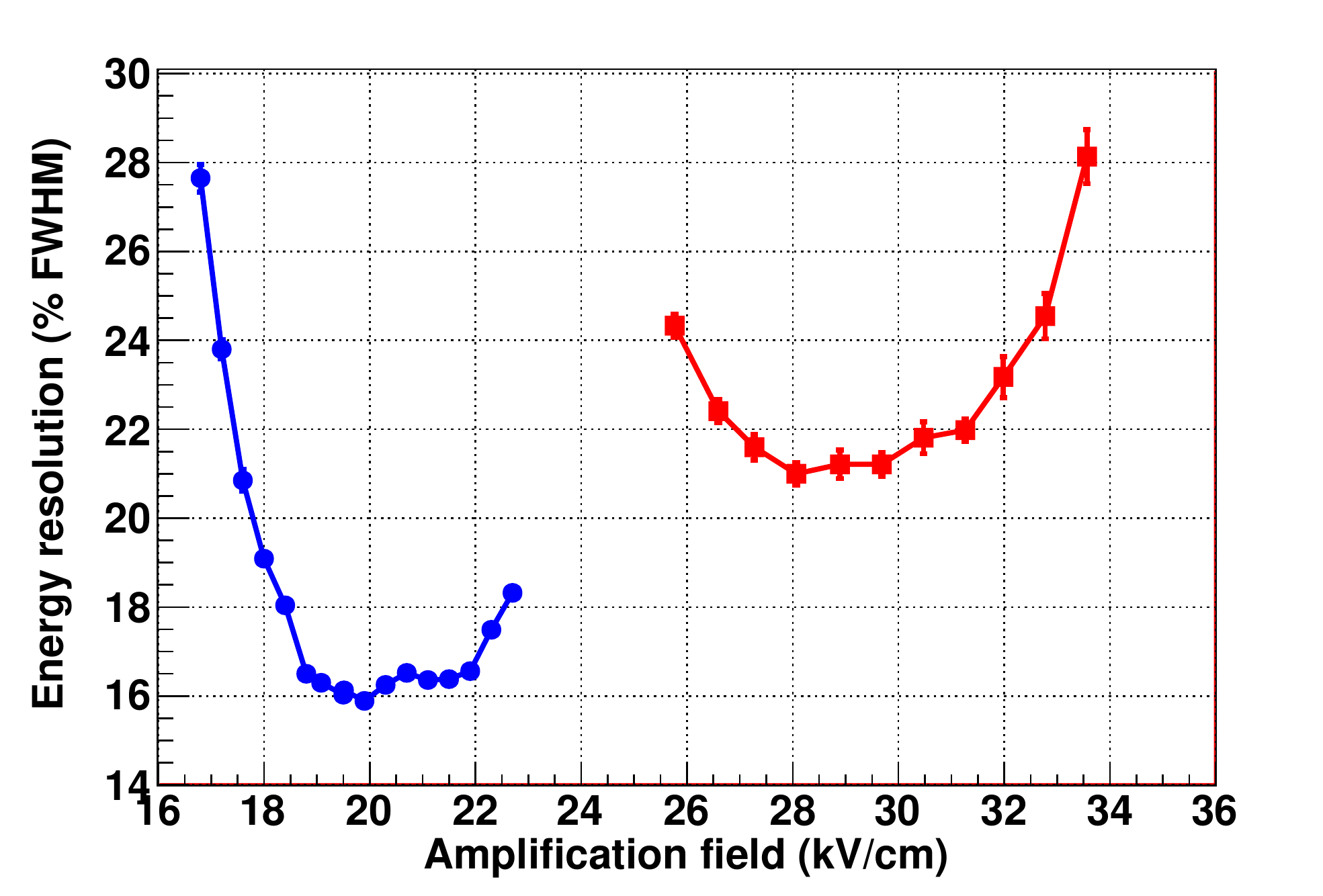}
\caption{Dependence of the energy resolution at 5.9 keV with the amplification field for the readouts of 128$\mu$m (red squared line) and 256 $\mu$m-thick amplification gaps (blue circle line).}
\label{fig:ERes10x10}
\end{figure}

As shown in figure \ref{fig:ERes10x10} for the best cases, the energy resolution stays constant for a wide range of amplification fields. At low fields, the resolution worsens due to the noise level that is comparable to the signal height. At high fields, it deteriorates due to the gain fluctuations. The values measured for each detector are shown in table \ref{tab:eres10x10}. We note that the readouts with a gap of 256\,$\mu$m show a better energy resolution than those with a gap of 128\,$\mu$m. This effect is  possibly due to the thickness of the bulk mesh. Indeed  for both amplification gaps the same mesh thickness (30\,$\mu$m) was used. Therefore the non-uniformity of the gap will have a bigger effect on the electric field of smaller gaps.

\begin{table}[htb!]
\begin{center}
$$
\begin{array}{ccc}
\hline
{\mbox{Detector}}&{\mbox{Num}}&{\mbox{Energy Resolution}}\\
{}&{}&{\mbox{(\% FWHM)}}\\
\hline
{\mbox{128}\,\mu\mbox{m}}&{1}&{21.0 \pm 0.3}\\
{}&{2}&{23.4 \pm 0.4}\\
{}&{3}&{23.2 \pm 0.4}\\
{\mbox{256}\,\mu\mbox{m}}&{1}&{16.0 \pm 0.1}\\
{}&{2}&{17.8 \pm 0.3}\\
\hline
\end{array}
$$
\end{center}
\caption{ Energy resolutions measured at 5.9 keV for the 3 different readouts of 128\,$\mu$m and for 2 at 256\,$\mu$m-thick amplification gap.}
\label{tab:eres10x10}
\end{table}

\section{Track measurements with alpha particles }\label{sec:results}

The T2K electronics\cite{T2Kelec1, T2Kelec2} has been used to read the signals induced in the strips to fully validate the concept of MIMAC readouts for the reconstruction of tracks. Eight flat cables connect the strips signals to two \emph{Front End Cards.} Each card is equipped with four ASIC chips, called AFTER, which digitize in 511 samples the signals of 72 channels, which are previously amplified and shaped. Finally, the data of each AFTER is sent by a \emph{Front End Mezzanine} (FEM) card to a DAQ card and subsequently to the computer for recording. As the external trigger mode of the T2K DAQ has been used, a trigger signal has been created feeding the bipolar output of the ORTEC VT120 amplifier/shaper into a FAN IN/OUT Lecroy 428F and subsequently into a NIM-TTL converter. Strips pulses have been sampled every 20\, ns and the peaking time has been fixed to 100\,ns. The dynamic range is of 120\,fC.

\begin{figure}[htb!]
\centering
\includegraphics[width=90mm]{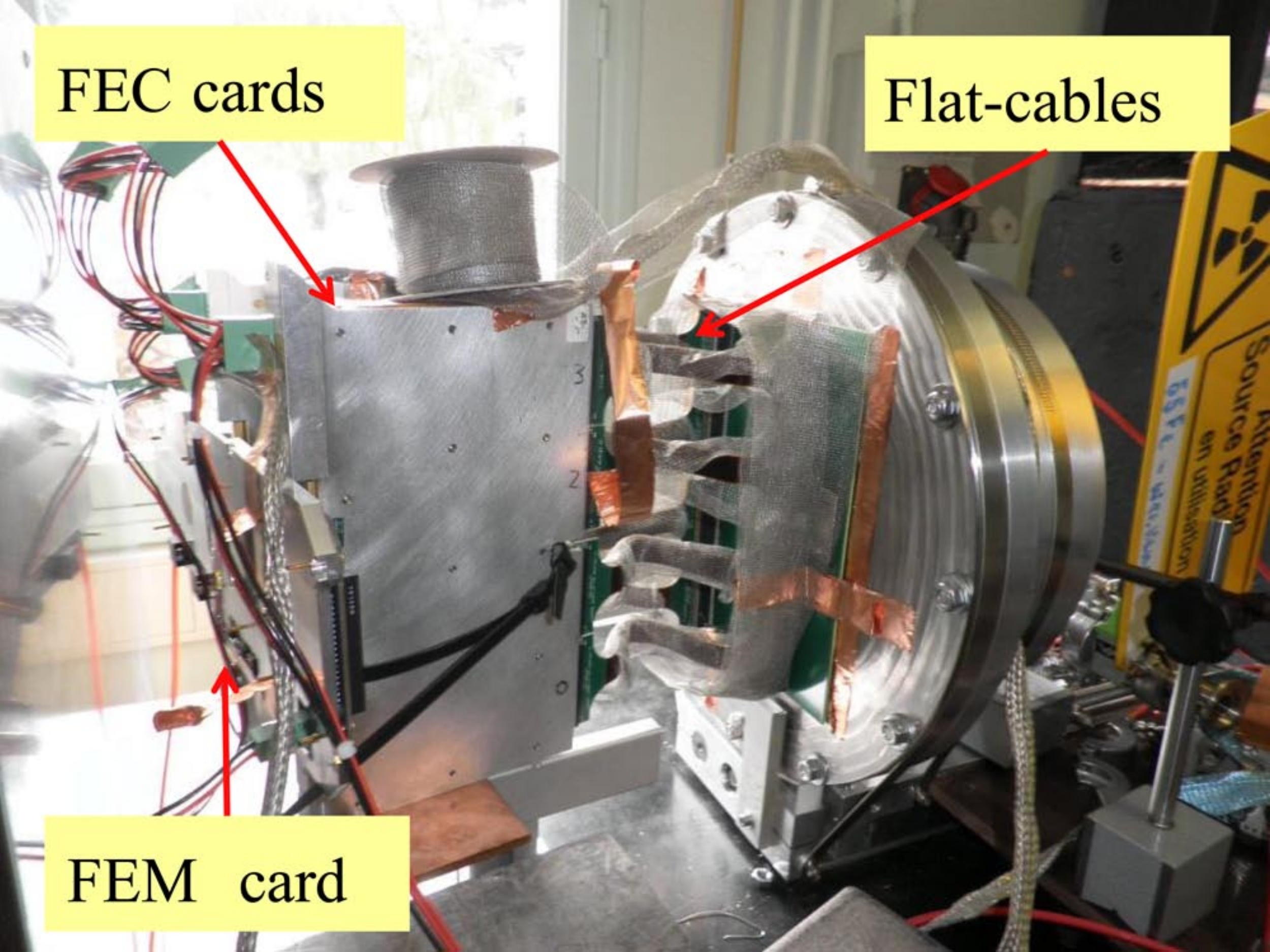}
\caption{A view of the dedicated vessel used to test MIMAC readouts when reading the strips with the T2K electronics. A detailed description is made in text.}
\label{fig:T2KElec}
\end{figure}

\begin{figure}[htb!]
\centering
\includegraphics[width=75mm]{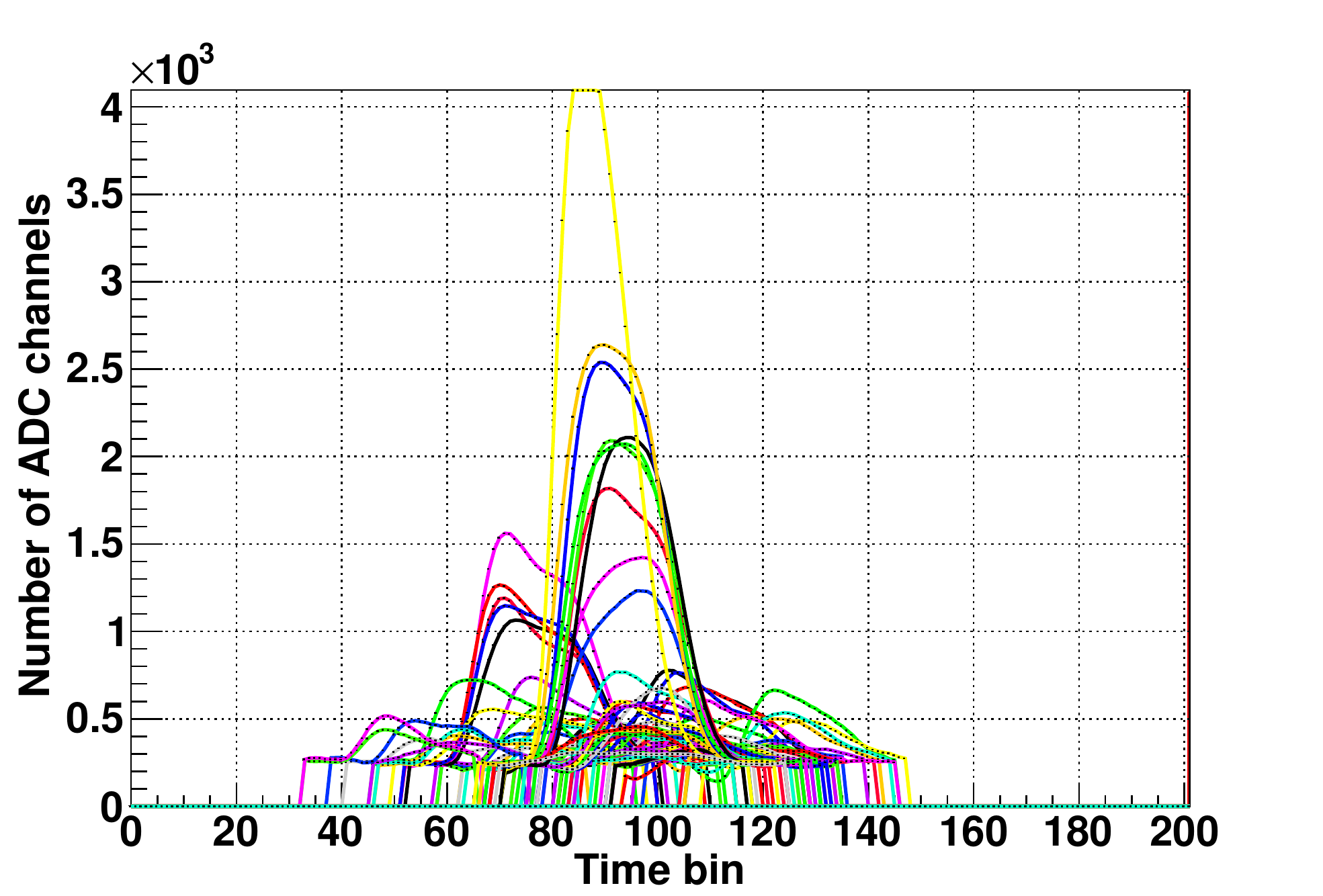}
\includegraphics[width=75mm]{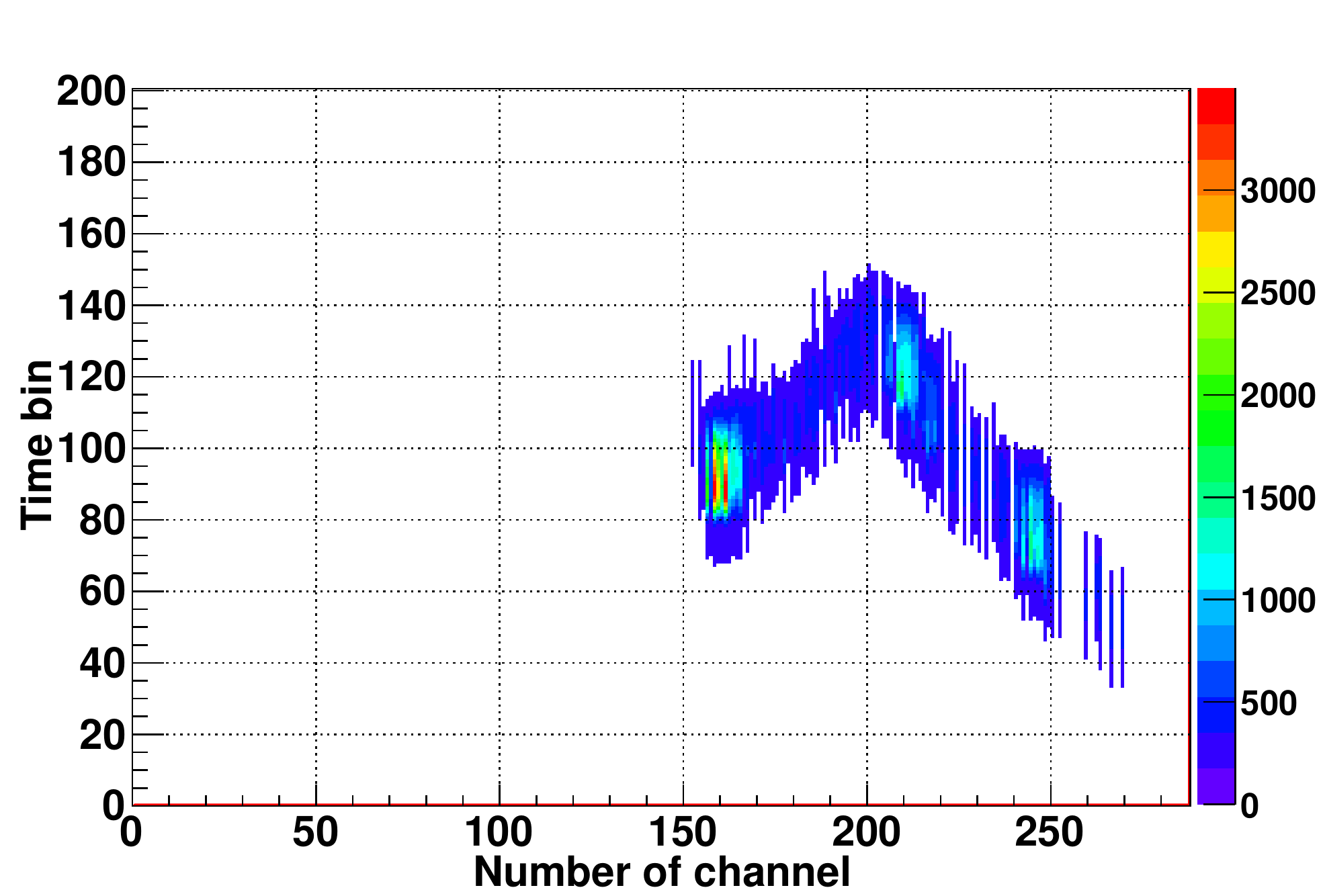}
\caption{ Left: Example of pulses induced in the strips acquired with the T2K electronics. Right: The reconstruction of the XZ projection of the same event. The physical event is an electron candidate of 44.4\,keV with a final charge accumulation (or blob).}
\label{fig:PulseEvent}
\end{figure}

In order to reconstruct the two 2D projection of each event, an
offline analysis software has been developped. It  extracts the strips
pulses by using the amplitude of each pulse sample and the readout
decoding. An example of the strips pulses and the XZ reconstruction of one event is shown in figure \ref{fig:PulseEvent}.

\begin{figure}[htb!]
\centering
\includegraphics[width=100mm]{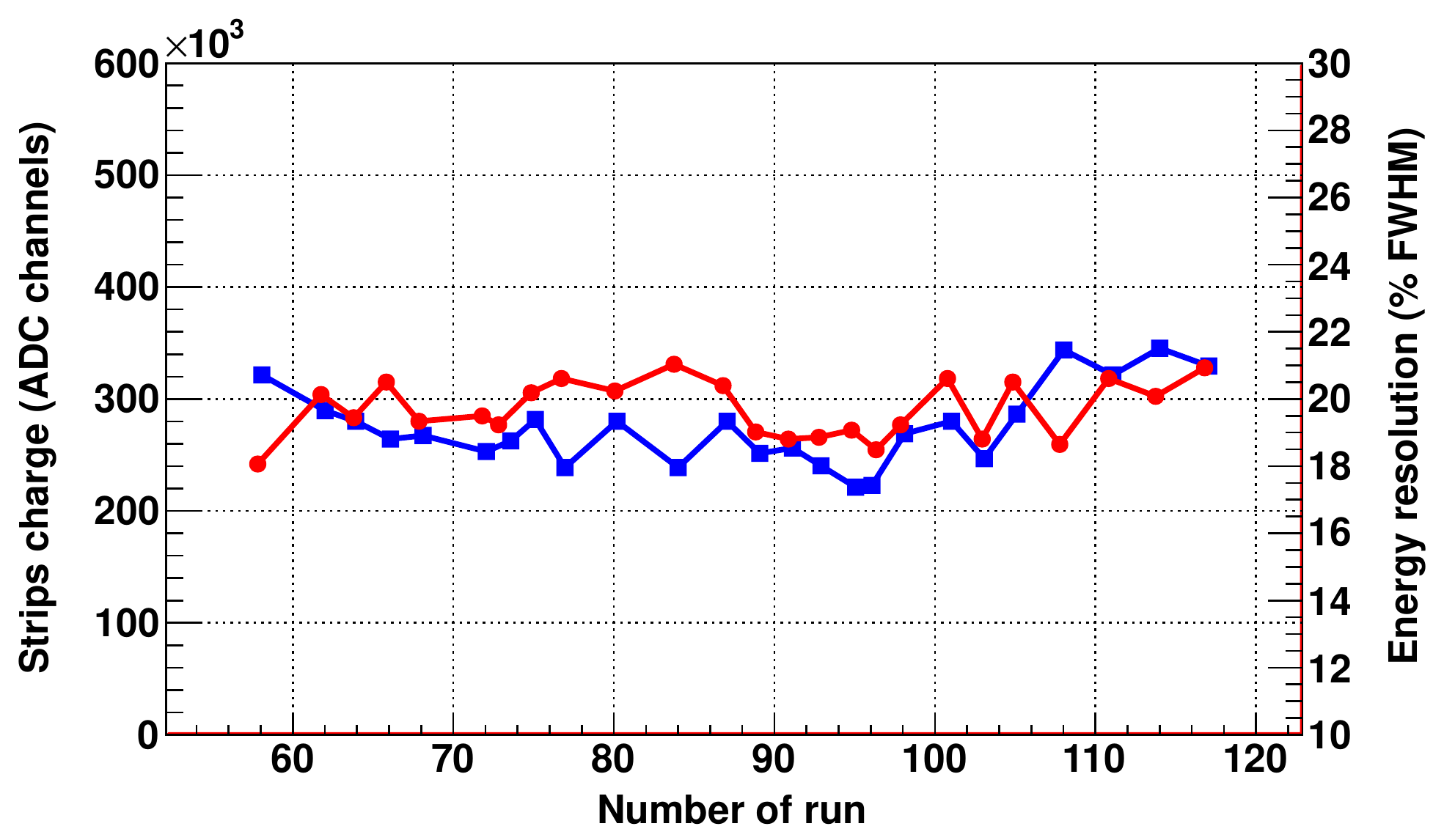}
\caption{ Evolution of the strips charge (red squared line) and the energy resolution at 5.9 keV (blue circled line) during the successive calibration runs with a source of $^{55}$Fe. The temporal distance between each calibration run is half day.} \label{fig:stability} \end{figure}

During three weeks of data-taking, a constant flow of 5 l/h of Ar+5\%iC$_4$H$_{10}$ was circulated in the dedicated vessel. The detector (with 256\,$\mu\mathrm{m}$ of amplification gap) was maintained in voltage 
($\mathrm{E_{amp}=21.9\,kV/cm}$, $\mathrm{E_{drift} = 88\,V/cm}$) acquiring events in a continous way. The detector was calibrated with an iron $^{55}$Fe source twice per day to monitor the evolution of the gain, energy resolution and the parameters calculated. The gain fluctuations observed were below 10\% due to the variations of pressure and temperature inside the vessel. It must be stressed that the pressure and the temperature were not controlled. The energy resolution varied between 18 and 20\% FWHM during the same period as can be seen in figure~\ref{fig:stability}.

In the two spatial projections (x and y) of the event, different parameters characterising the charge were calculated like mean position, width and number of activated strips. The analysis was then extended to the perpendicular direction using the amplitudes of strips pulse in each temporal bin. Finally, the total charge of each event was obtained summing the charge of both projections. 
After having tested the readout with low energy events, we evaluated its performance at low gain with high energy events. For this purpose, an $^{241}$Am alpha source was installed in the source keeper screwed at the center of the drift plate. The source consists of a metallic circular substrate of 25\,mm diameter where the radioactive material has been deposited on its center in a circular region of about 8\,mm of diameter. The alpha particles are emitted isotropically. The 5.5 MeV\,alphas were used to characterize the readout as it was done with the iron source in section \ref{sec:characterisation}, generating the electron transmission and gain curves. The electron transmission curve, taken at an absolute gain of 85, matched with the one showed in figure \ref{fig:Trans10x10}. Meanwhile, the gain curve was produced varying the mesh voltage between 270 and 420\,V, as shown in figure \ref{fig:Gain10x10alpha}, and follows the tendency of the one generated by photons.

\begin{figure}[htb!]
\centering
\includegraphics[width=75mm]{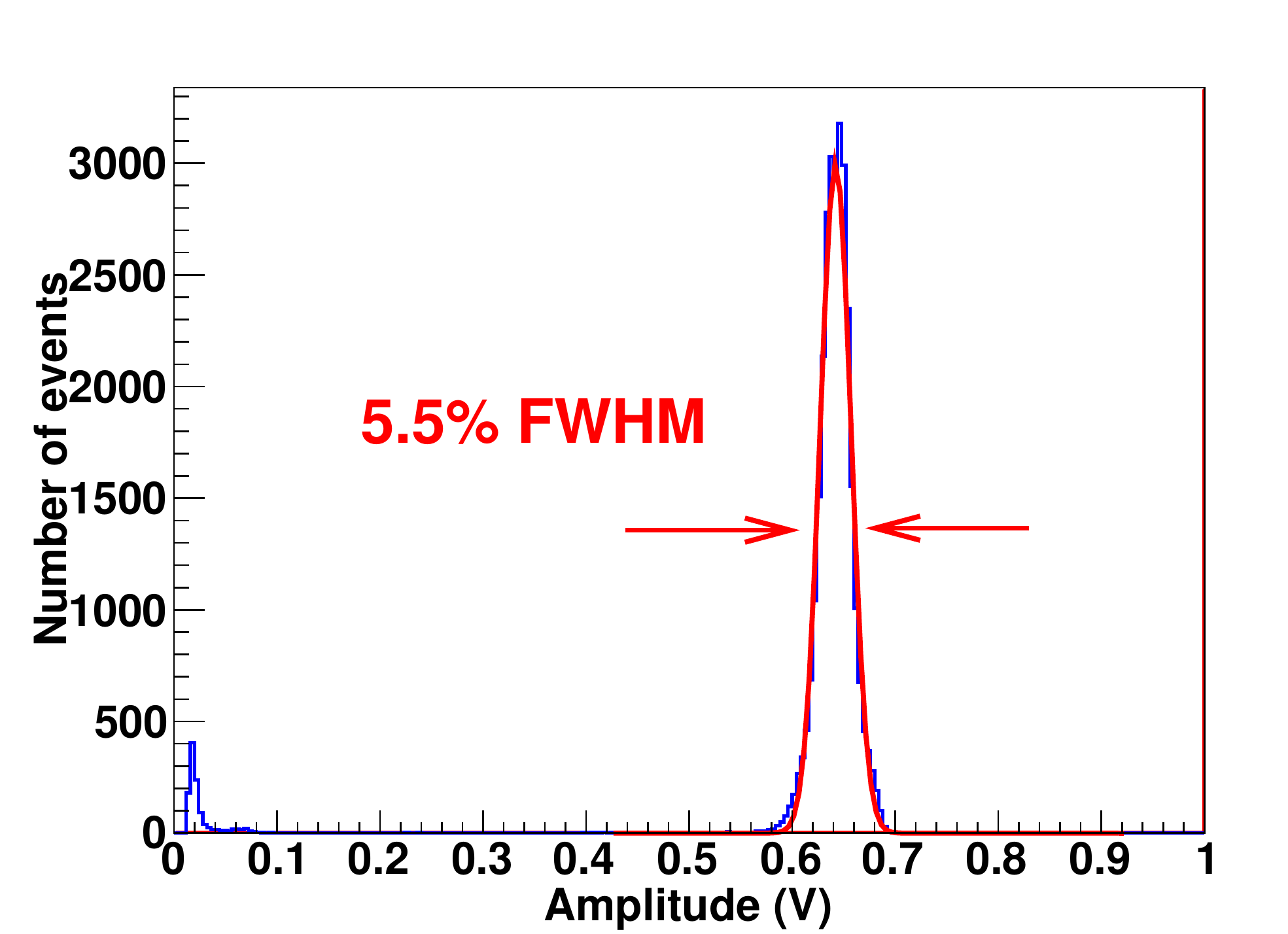}
\includegraphics[width=75mm]{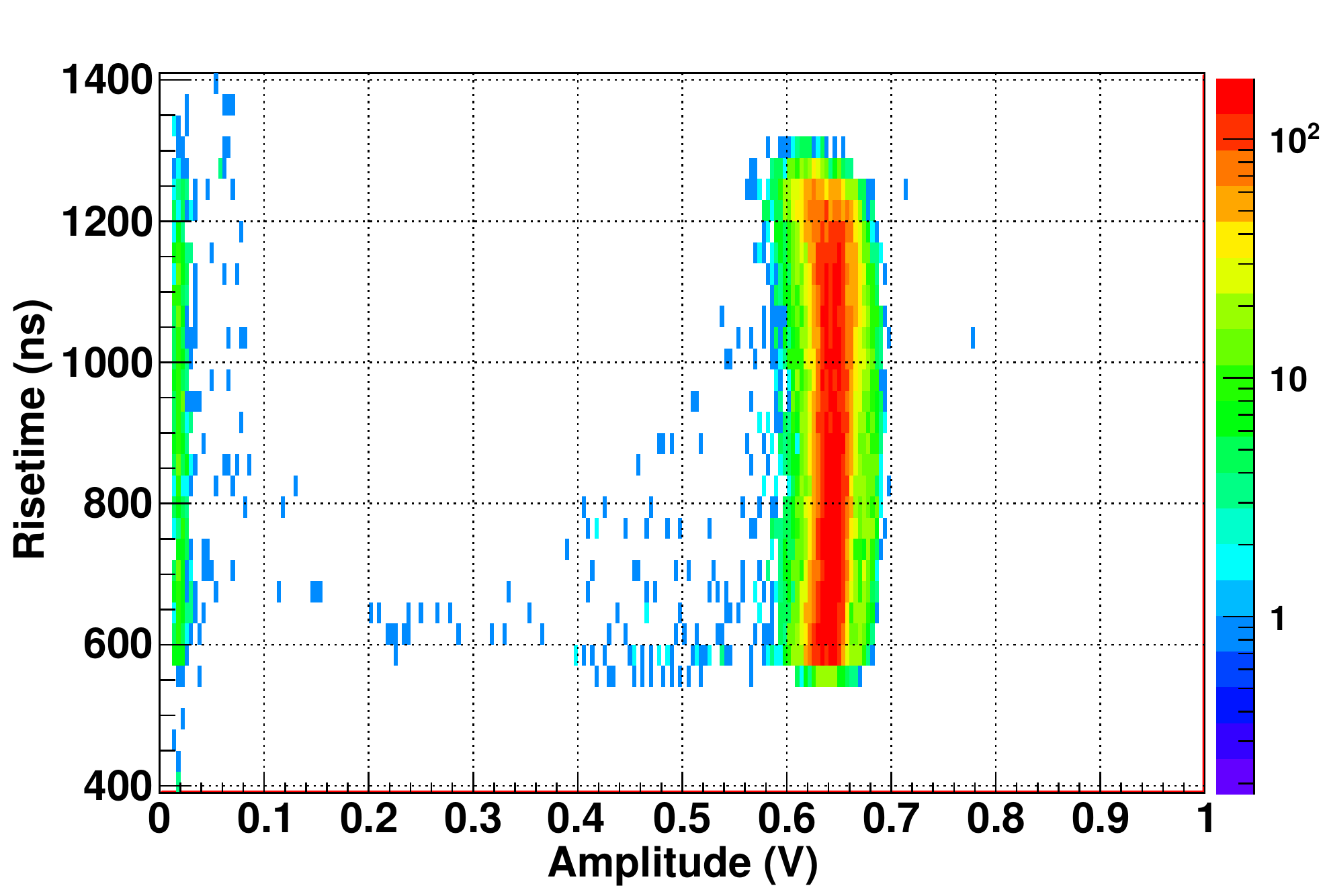}
\caption{ Left: Spectrum generated by the mesh pulses induced by the $^{241}$Am alphas, showing an energy resolution of 5.5\% FWHM. Right: Distribution of the risetime versus the amplitude of the mesh pulses induced by the $^{241}$Am alphas.}
\label{fig:SpecAmpRise}
\end{figure}

The spectra generated by the $^{241}$Am source showed an energy resolution of 5.5\% FWHM, as the one shown in figure \ref{fig:SpecAmpRise}. This value was independent of the drift voltage and the readout gain. To check the possible presence of attachment effects in the gas, mesh pulses were acquired by a LeCroy WR6050 oscilloscope. In an offline analysis, the amplitude and risetime of the pulses were calculated and the 2D distribution of these parameters was generated to look for correlations. Alpha events showed the same amplitude, independently of their risetime, i.e., their spatial direction. Therefore attachement effects are not observed. An example of these 2D distribution is shown in figure \ref{fig:SpecAmpRise} (right).

\begin{figure}[htb!]
\centering
\includegraphics[width=75mm]{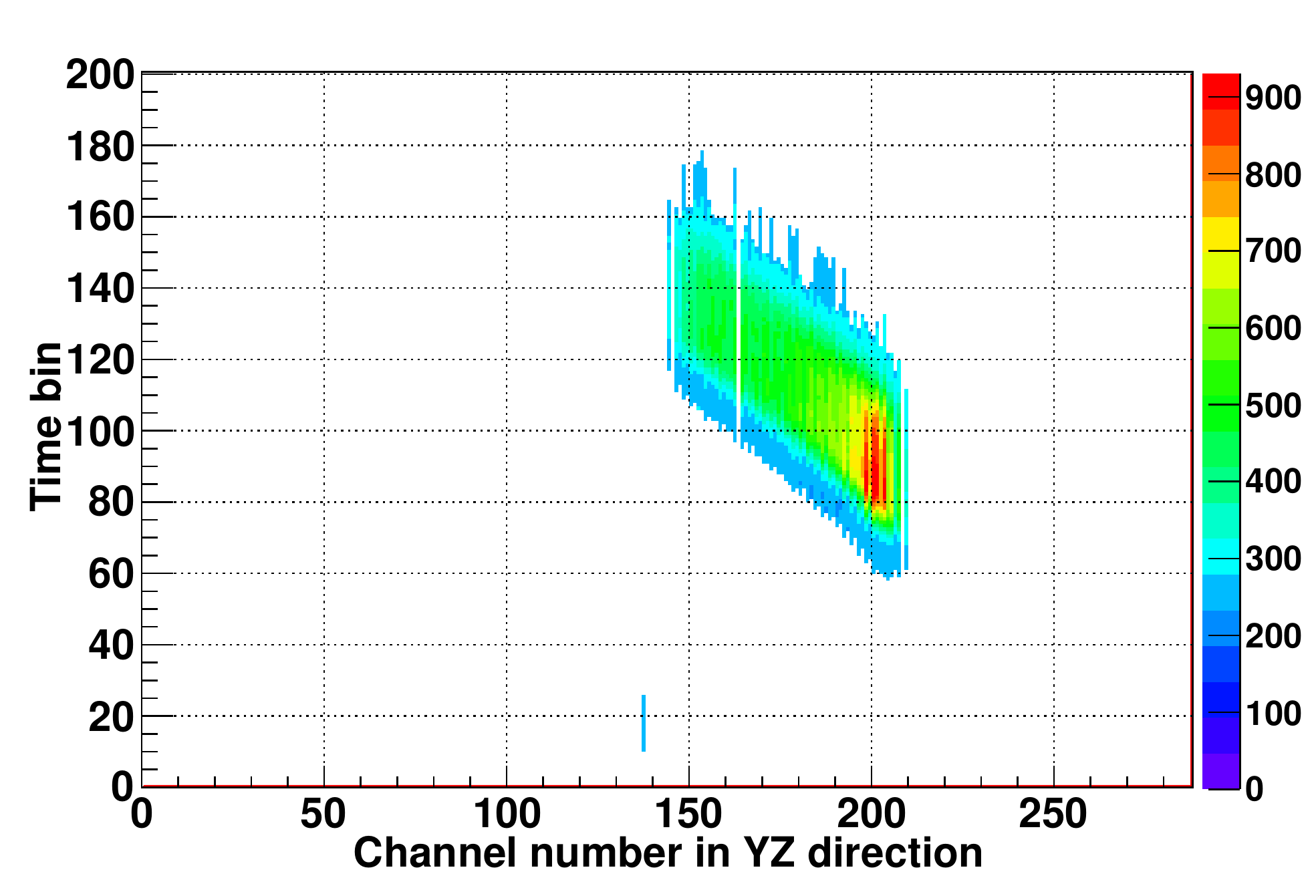}
\includegraphics[width=75mm]{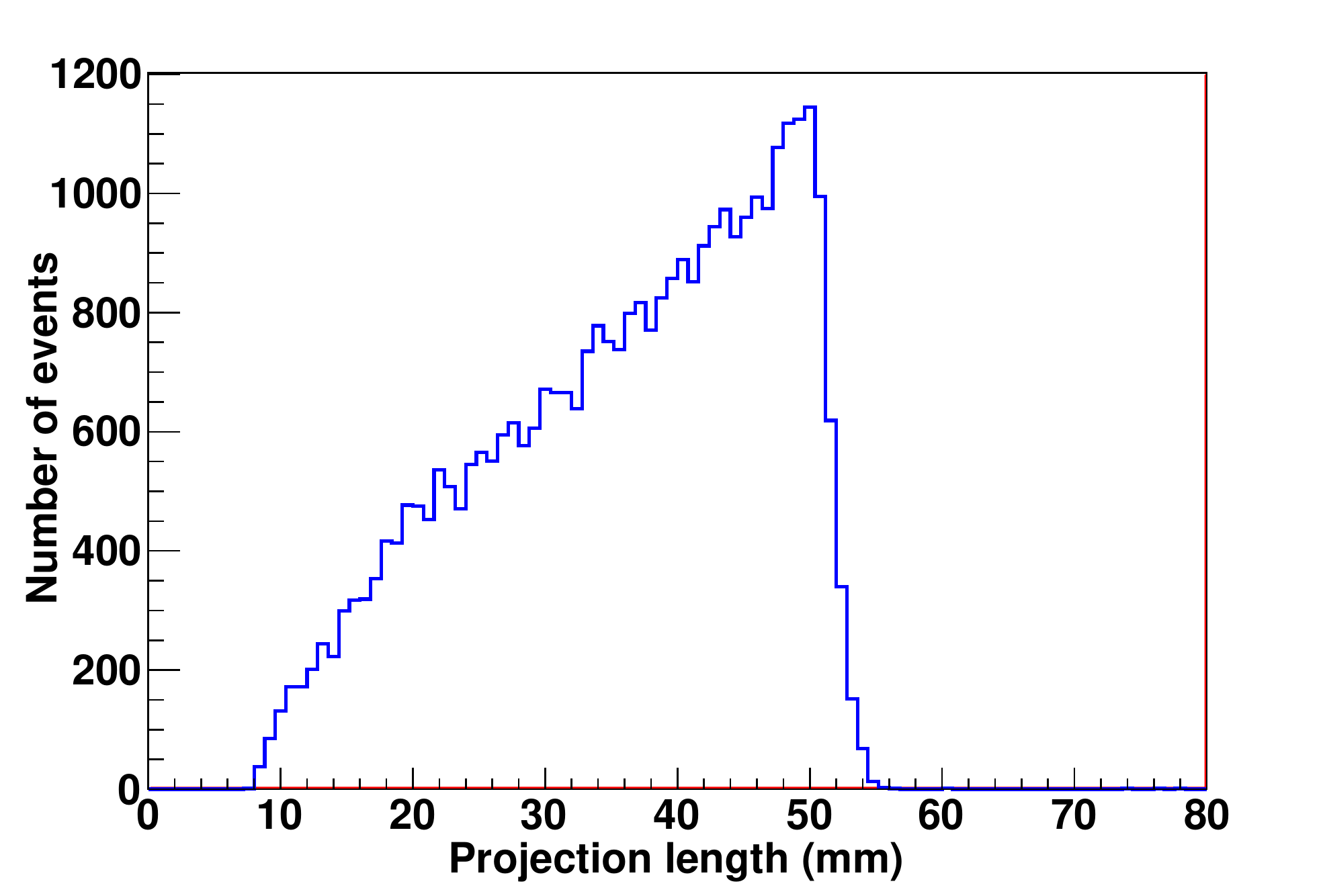}
\caption{ Left: An alpha event acquired by the T2K electronics. The track shows a final charge accumulation. Right: Distribution of the length of the track projection on the XY plane, calculated by the number of strips activated in each direction. The maximum length of the distribution matches the Geant4 simulation value (54\,mm).}
\label{fig:ProjLeng}
\end{figure}

Several alpha tracks were also acquired with the T2K electronics, as shown in figure \ref{fig:ProjLeng} (left). The mesh and drift voltages were respectively set to 400 and 820\,V, which correspond to a gain of 85 and a drift field of 70~V/cm. For each event, the length of the track projection on the XY plane was calculated and the distribution of this variable was generated. The maximum value obtained (54\,mm) matches the theoretical length expected for a 5.5 MeV\,alpha particles in an argon-based mixture\cite{IguazphD}.

\section{Conclusions and Perspectives}
\label{sec:conclusions}
The readout plane of the Bulk 10 $\times$ 10\,cm$^2$ Micromegas designed and built for the MIMAC project has been described. The first characterisation tests in the laboratory show good performance in terms of gain, uniformity, energy resolution and track measurements. The following steps are to test this detector in a neutron beam facility with neutrons of few keV using the specifically designed MIMAC electronics to reach the ultimate performance of the detector for the detection of Dark Matter. Plans for a test in the Frejus underground laboratory with a module containing two 10 $\times$ 10\,cm$^2$ detectors are envisaged in the next coming months. This measurement, in a realistic environment for a Dark Matter experiment, will give crucial information (background rejection as well as intrinsic contamination of the used materials) before the construction of a 1\,m$^3$ experiment. In any case the results obtained up to now validate the MIMAC concept for the construction of a large TPC for directional detection of Dark Matter.

\acknowledgments
The MIMAC Collaboration acknowledges the ANR-07-BLANC-0255-03 funding. The authors would like to thank D. Desforges for his availability in the use of the Mitutuyo Microscope as well as D. Jourde for his help with the  degrador.

\end{document}